\documentclass[11pt]{article}

\usepackage[bookmarks=true]{hyperref}
\hypersetup{breaklinks=true} 

\let\fn\footnote
\renewcommand{\footnote}[1]{\linespread{1.1}\fn{#1}\linespread{1.29}}

\usepackage{fancyhdr}
\usepackage[left]{lineno}

\makeatletter\renewcommand{\section}{\@startsection
{section}{1}{\z@}{-3.5ex plus -1ex minus
    -.2ex}{2.3ex plus .2ex}{\bf }}
\makeatletter\renewcommand{\subsection}{\@startsection{subsection}{2}{\z@}{-3.25ex
plus -1ex minus
   -.2ex}{1.5ex plus .2ex}{\it }}
\makeatletter\renewcommand{\subsubsection}{\@startsection{subsubsection}{3}{-2.45ex}{-3.25ex
plus -1ex minus -.2ex}{1.5ex plus .2ex}{\it }}
\renewcommand{\thesection}{\arabic{section}.}
\renewcommand{\thesubsection}{\arabic{section}.\arabic{subsection}.}

\renewcommand{\theequation}{\thesection\arabic{equation}}
\makeatletter \@addtoreset{equation}{section}

\renewenvironment{thebibliography}[1]
     {\baselineskip=16pt plus 2pt minus 1pt
      \section*{\large\refname
        \@mkboth{\MakeUppercase\refname}{\MakeUppercase\refname}}%
     \list{\@biblabel{\@arabic\c@enumiv}}%
           {\settowidth\labelwidth{\@biblabel{#1}}%
            \leftmargin\labelwidth
            \advance\leftmargin\labelsep
            \@openbib@code
            \usecounter{enumiv}%
            \let\p@enumiv\@empty
            \renewcommand\theenumiv{\@arabic\c@enumiv}}%
      \sloppy
      \clubpenalty4000
      \@clubpenalty \clubpenalty
      \widowpenalty4000%
      \sfcode`\.\@m}

\newcommand{\acknowledgements}{\section*{Acknowledgements}
\addcontentsline{toc}{section}{\hspace{0.6cm}{\bf Acknowledgements}}}

\newcommand{\appendices}{\section*{Appendix}
\setcounter{subsection}{0}\setcounter{equation}{0}\renewcommand{\thesubsection}{\Alph{subsection}.}
\renewcommand{\theequation}{\thesubsection\arabic{equation}}
\addcontentsline{toc}{section}{\hspace{0.6cm}{\bf Appendix}}
}

\usepackage{epsfig,rotating}
\usepackage{amsmath,amssymb}
\usepackage{amsfonts}
\usepackage{mathrsfs}
\usepackage{bbm}
\usepackage{bm}

\renewcommand{\slash}[1]{#1\hspace{-0.27cm}/\,}
\newcommand{\nablas}{\slash{\nabla}}
\newcommand{\nablabs}{\slash{\bar{\nabla}}}

\def\periodb#1{\setbox0=\hbox{$#1$}#1\hskip-\wd0\hbox to\wd0{-}}

\newcommand{\unit}{\mathbbm{1}}   			

\newcommand{\Cv}{\check{C}}
\newcommand{\Hv}{\check{H}}

\newcommand{\CA}{\mathcal{A}}    			

\newcommand{\CC}{\mathcal{C}}

\newcommand{\CI}{\mathcal{I}}

\newcommand{\CL}{\mathcal{L}}

\newcommand{\CN}{\mathcal{N}}
\newcommand{\CO}{\mathcal{O}}

\newcommand{\CP}{\mathcal{P}}

\newcommand{\CT}{\mathcal{T}}

\newcommand{\CU}{\mathcal{U}}

\newcommand{\CE}{\mathcal{E}}
\newcommand{\frg}{\mathfrak{g}}				

\newcommand{\FR}{\mathbbm{R}}     			
\newcommand{\FC}{\mathbbm{C}}     			
\newcommand{\RZ}{\mathbbm{Z}}     			

\newcommand{\xd}{\dot{x}}
\newcommand{\dd}{\mathrm{d}}     			
\newcommand{\dpar}{\partial}     			
\newcommand{\de}{\mathrm{e}}     			
\newcommand{\di}{\mathrm{i}}     			
\newcommand{\eps}{{\varepsilon}}			

\newcommand{\psib}{{\bar{\psi}}}

\newcommand{\eand}{{~~~\mbox{and}~~~}}     		
\newcommand{\ewith}{{~~~\mbox{with}~~~}}

\newcommand{\eor}{{~~~\mbox{or}~~~}}

\newcommand{\der}[1]{\frac{\dpar}{\dpar #1}}   		
\newcommand{\derdel}[1]{\frac{\delta}{\delta #1}}   		
\newcommand{\dder}[1]{\frac{\dd}{\dd #1}}   		
\newcommand{\dderr}[2]{\frac{\dd #1}{\dd #2}}   	

\newcommand{\au}{\mathfrak{u}}
\newcommand{\asu}{\mathfrak{su}}

\newcommand{\sU}{\mathsf{U}}     			

\newcommand{\sSU}{\mathsf{SU}}

\newcommand{\sSO}{\mathsf{SO}}

\newcommand{\acton}{\vartriangleright}     			
\newcommand{\remark}[1]{}     				
\def\tyng(#1){\hbox{\tiny$\yng(#1)$}}			
\def\tyoung(#1){\hbox{\tiny$\young(#1)$}}			

\begin{document}
\begin{titlepage}
\begin{flushright}
 HWM--10--25 \\ EMPG--10--12
\end{flushright}
\vskip 2.0cm
\begin{center}
{\LARGE \bf Constructing Self-Dual Strings}
\vskip 1.5cm
{\Large Christian S{\"a}mann}
\setcounter{footnote}{0}
\renewcommand{\thefootnote}{\arabic{thefootnote}}
\vskip 1cm
{\em Department of Mathematics\\
Heriot-Watt University\\
Colin Maclaurin Building, Riccarton, Edinburgh EH14 4AS, U.K.\\
and Maxwell Institute for Mathematical Sciences, Edinburgh,
U.K.}\\[5mm]
{Email: {\ttfamily C.Saemann@hw.ac.uk}} \vskip
1.1cm
\end{center}
\vskip 1.0cm
\begin{center}
{\bf Abstract}
\end{center}
\begin{quote}
We present an ADHMN-like construction which generates self-dual string solutions to the effective M5-brane worldvolume theory from solutions to the Basu-Harvey equation. Our construction finds a natural interpretation in terms of gerbes, which we develop in some detail. We also comment on a possible extension to stacks of multiple M5-branes.
\end{quote}
\end{titlepage}

\section{Introduction}

There is a close link between certain supersymmetric D-brane configurations and classical integrability as found in the self-dual Yang-Mills equation and its dimensional reductions. For example, the Atiyah-Drinfeld-Hitchin-Manin (ADHM) construction of instantons \cite{Atiyah:1978ri} finds a full interpretation within the gauge theoretic description of D0-D4-brane bound states, see e.g.\ \cite{Tong:2005un} for a review. Similarly, Nahm's extension \cite{Nahm:1979yw,Nahm:1981nb,Nahm:1982jb} of the ADHM construction to the case of monopoles, the ADHMN construction, is reflected in the effective description of D$1$-branes ending on D$3$-branes \cite{Diaconescu:1996rk,Tsimpis:1998zh}. In both constructions, one starts from a Dirac operator and constructs the instanton and monopole solutions from its zero modes in a straightforward manner. The Dirac operator in turn is fully determined by solutions to a matrix equation in the ADHM case and solutions to the Nahm equation, i.e.\ the dimensional reduction of the self-dual Yang-Mills equation to one dimension, in the ADHMN case.

It is clearly interesting to look for such a link between integrable field theories and brane configurations also in M-theory. The obvious starting point here is a configuration of M2-branes ending on an M5-brane, which is obtained from the D1-D3-brane configuration describing monopoles via T-duality and taking the M-theory limit. For this configuration, Basu and Harvey \cite{Basu:2004ed} suggested a new Nahm-type equation, in which the Lie algebra structure is replaced by a 3-Lie algebra\footnote{Basu and Harvey actually suggested a matrix algebra closely related to the 3-Lie algebra $A_4$. The usefulness of general 3-Lie algebras in this context was first observed in \cite{Bagger:2006sk}.}. This Basu-Harvey equation passed many consistency tests and in particular, it led to the remarkably successful Bagger-Lambert-Gustavsson (BLG) model \cite{Bagger:2007jr,Gustavsson:2007vu} and its close relative, the Aharony-Bergman-Jafferis-Maldacena (ABJM) model \cite{Aharony:2008ug}. We therefore assume that the Basu-Harvey equation is indeed the appropriate substitute for the Nahm equation. 

In this paper, we propose a completion of the picture: We present a Dirac operator built from solutions to the Basu-Harvey equation and demonstrate how its zero modes can be used to construct solutions to the self-dual string equation, which is the analogue of the Bogomolny monopole equation in the case of the M2-M5-brane configuration. We hope that this construction can help to understand the structures which underlie an effective description of multiple M5-branes.

Note that an earlier approach to an ADHMN construction for self-dual strings was suggested by Gustavsson \cite{Gustavsson:2008dy}. There, the author worked on an auxiliary space obtained from the loop space $\CL\FR^4$, described by the coordinates $\sigma^{\mu\nu}(\tau):=x^\mu(\tau)\xd^\nu(\tau)-x^\nu(\tau)\xd^\mu(\tau)$ where $x^\mu(\tau)\in\CL\FR^4$. He then split the auxiliary space up into two by grouping these six coordinates into pairs of three using the 't Hooft tensors. Eventually, he proposed to identify self-dual strings with a pair of monopoles living in each of these two new auxiliary spaces. Our construction, however, is different: We start from a gerbe over $S^3$, which we transgress, together with the self-dual string equation, to a principal $\sU(1)$-bundle over the loop space $\CL S^3$. The resulting self-dual string equation is different from the pair of Nahm equations used in \cite{Gustavsson:2008dy}.

Throughout the paper, we shall try to be mostly self-contained and to give detailed motivation for our constructions. In section 2, we present some mathematical background on gerbes before we begin our discussion in section 3 by reviewing the standard ADHMN construction. Section 4 contains the discussion of our extension of this construction to self-dual strings and we conclude in section 5. An appendix reviews metric 3-Lie algebras and introduces the notion of compatible representations of their associated Lie algebras.

\section{Gerbes with connective structure}

In the following, we review the necessary definitions of a Hitchin-Chatterjee gerbe with connection; readers familiar with this notion can skip this section. We will focus on gerbes over $S^3$, as these will motivate our generalization of the ADHMN construction. Our discussion follows mainly \cite{Murray:0712.1651} and \cite{Hitchin:9907034}, see also  \cite{0817647309} for a standard reference on gerbes as well as \cite{Chatterjee:1998}. We also found the webpages of nLab\footnote{\href{http://ncatlab.org/nlab/show/HomePage}{{\tt http://ncatlab.org/nlab/show/HomePage}}} to be helpful.

\subsection[From U(1)-bundles to gerbes]{From $\sU(1)$-bundles to gerbes}

Consider a principal $\sU(1)$-bundle $P$ over a manifold $M$ and let $\CU=\{U_i\}$ be a cover of $M$. The bundle $P$ can be defined by transition functions which are \v{C}ech cocycles\footnote{The dependence on the cover is removed as usual by taking direct limits.} $[g_{ij}]\in \Hv^1(\CU,\CC^\infty(S^1))$. We can now introduce a connection $\nabla$ on $P$ by specifying a \v{C}ech 1-cochain of $\au(1)$-valued one-forms $A_i$ such that on overlaps $U_i\cap U_j$, we have $A_i-A_j=\dd \log g_{ij}$. As $\dd A_i-\dd A_j=0$, the curvature $F:=\dd A$ of the connection $\nabla$ is globally defined. It forms a representative of the Chern class of $P$ and we have $F\in H^2(M,\RZ)$. Conversely, given a curvature two-form $F\in H^2(M,\RZ)$, repeated application of the Poincar{\'e} lemma yields a transition function on any overlap $U_i\cap U_j$ representing the class $[g_{ij}]$.

Note also that $\Hv^1(M,\CC^\infty(S^1))\cong H^2(M,\RZ)$, which follows from the long exact cohomology sequence of the short exact sequence
\begin{equation}
 0\longrightarrow \RZ\longrightarrow \CC^\infty(\FR)\stackrel{\de^{2\pi\di x}}{\longrightarrow} \CC^\infty(S^1)\longrightarrow 1~~,
\end{equation}
the fact that $\CC^\infty(\FR)$ is a fine sheaf (implying $H^i(M,\CC^\infty(\FR))=0$) and the standard isomorphism between de Rham and \v{C}ech cohomology groups.

Gerbes over manifolds are generalizations of $\sU(1)$-bundles in that they correspond to \v{C}ech cocycles $[g_{ijk}]\in \Hv^2(M,\CC^\infty(S^1))$ or equivalently to elements of the cohomology class $H^3(M,\RZ)$.

\subsection{Local gerbes}

It will be sufficient for us to work with surjective submersions obtained from open covers\footnote{This means that we restrict ourselves to {\em local} or {\em Hitchin-Chatterjee gerbes}.} of manifolds. That is, if $\CU=(U_i)$ is an open cover of a manifold, we consider the associated disjoint union of patches,
\begin{equation}
 Y_\CU:=\{(i,x)|x\in U_i\}~,
\end{equation}
together with the surjective submersion $\pi:Y_\CU\twoheadrightarrow M$ with $\pi(i,x):=x$. We will also need the ordered, $p$-fold intersections of (not necessarily distinct) patches
\begin{equation}
 Y^{[p]}_\CU:=\{(y_1,\ldots,y_p)|\pi(y_1)=\ldots=\pi(y_p)\}\subset Y^p_\CU~.
\end{equation}
We denote by $\pi_i:Y^{[p]}_\CU\rightarrow Y^{[p-1]}_\CU$ the obvious projection by dropping the $i$th patch. Note that on the $Y^{[p]}_\CU$, we have the usual \v{C}ech de Rham double complex
\begin{equation}
 0\rightarrow \Omega^q(M)\stackrel{\pi^*}{\longrightarrow}\Omega^q(Y_\CU)\stackrel{\delta}{\longrightarrow}\Omega^q(Y^{[2]}_\CU)\stackrel{\delta}{\longrightarrow}\ldots~,
\end{equation}
where $\delta=\sum_{i=1}^p(-1)^{p-1}\pi_i^*$ is the standard \v{C}ech differential (in additive notation). Although we are working with the abelian sheaf $\CC^\infty(S^1)$, we switch to multiplicative notation in the following.  

A {\em local bundle gerbe} over a manifold $M$ is given \cite{Murray:0712.1651} by a pair $(P,Y_\CU)$, where $Y_\CU\twoheadrightarrow M$ is a subjective submersion obtained from a cover $\CU$ and $P\rightarrow Y^{[2]}_\CU$ is a $\sU(1)$-bundle, together with a compatible {\em bundle gerbe multiplication}. The latter can be seen as the analogue of the tensor product of $\sU(1)$-bundles.

The bundle gerbe multiplication $\mu$ yields the following smooth isomorphism of $\sU(1)$-bundles over $Y^{[3]}$:
\begin{equation}
 \mu: \pi^*_3(P)\otimes\pi^*_1(P)\rightarrow \pi^*_2(P)~.
\end{equation}
Given an element on $Y^{[2]}_\CU$ as $(i,j,x)$, where $x\in U_i\cap U_j$, we have explicitly
\begin{equation}
 \mu:((i,j,x),z_1)\otimes((j,k,x),z_2)\mapsto((i,k,x),z_1z_2g_{ijk}(x))
\end{equation}
for some $g_{ijk}\in \Cv^2(\CU,\CC^\infty(S^1))$. We now demand that this multiplication is associative in the following sense: Denote by $P_{y_1,y_2}$ the fibre of $P$ over the point $(y_1,y_2)$. Then the diagram 
\begin{equation}
\begin{array} {ccc}
 P_{y_1,y_2}\otimes P_{y_2,y_3}\otimes P_{y_3,y_4} & \longrightarrow & P_{y_1,y_3}\otimes P_{y_3,y_4}\\
 \downarrow & &\downarrow \\
P_{y_1,y_2}\otimes P_{y_2,y_4} & \longrightarrow & P_{y_1,y_4} 
\end{array}
\end{equation}
commutes for any $(y_1,y_2,y_3,y_4)\in Y^{[4]}$. This is the case exactly if $\delta(g)=1$, i.e.\ $g$ is a cocycle, as one readily verifies. The r{\^o}le the first Chern class plays for $\sU(1)$-bundles is taken over by the {\em Dixmier-Douady class} for gerbes, which corresponds to $\Hv^2(M,\CC^\infty(S^1))\cong H^3(M,\RZ)$.

\subsection{Connective structures on gerbes}

Consider a principal $\sU(1)$-bundle with connection over a manifold $M$ with open cover $\CU=(U_i)$. Its curvature $F$ is a globally defined 2-form, the connection corresponds to a $\au(1)$-valued one-form on each patch $U_i$ and on overlaps $U_i\cap U_j$, we have transition functions $f_{ij}$. The relations between these are
\begin{equation}
 F=\dd A_i~\mbox{on}~U_i\eand A_i-A_j=\dd \log f_{ij}~\mbox{on}~U_i\cap U_j~.
\end{equation}

On gerbes, we shift these objects by one form degree or, equivalently, by one degree in their \v{C}ech cohomology. That is, we have a global curvature 3-form $H$, two-forms $B_i$ on the patches $U_i$, one-forms $A_{ij}$ on the intersections $U_i\cap U_j$ and functions $h_{ijk}$ on triple intersections $U_i\cap U_j\cap U_k$, all taking values in $\au(1)$. Up to obvious equivalences, we can start from $H$ and construct the other objects by the Poincar{\'e} lemma, trading form degree for \v{C}ech degree. Explicitly, the relations between these are 
\begin{equation}
\begin{aligned}
 H=\dd B_i~\mbox{on}~U_i~,~~~B_i-B_j=\dd A_{ij}~\mbox{on}~U_i\cap U_j~,\\
A_{ij}-A_{ik}+A_{jk}=\dd h_{ijk}~\mbox{on}~U_i\cap U_j\cap U_k~.~~~~
\end{aligned}
\end{equation}
If $H\in H^3(M,\RZ)$, then $g_{ijk}:=\exp(\di h_{ijk})$ is a \v{C}ech cocycle. The one-forms $A_{ij}$ are interpreted as giving rise to connections on the $\sU(1)$-bundle over $Y_\CU^{[2]}$ defining the gerbe and $A$ is called a {\em bundle gerbe connection}. The two-forms $B_i$ are called a {\em curving} for $A$, and the combined data $(A,B)$ yields a {\em connective structure} on the gerbe $(P,Y_\CU)$.

\subsection[Gerbes over S^^3]{Gerbes over $S^3$}

Consider now $M=S^3$ and let the sphere be covered by two patches $\CU=\{U_0,U_1\}$ containing the north and the south pole, respectively. The intersection $U_0\cap U_1$ can be identified with the space $S^2\times (-1,1)$. As a gerbe is primarily defined by a $\sU(1)$-bundle over $Y^{[2]}_\CU$, we see that gerbes over $S^3$ correspond to $\sU(1)$-bundles over $S^2$. At cohomological level, this is reflected in $H^2(S^2,\RZ)\cong H^3(S^3,\RZ)\cong\RZ$. We can pull back the whole $\sU(1)$-bundle over $S^2$ including its connection $A$ and curvature $F$. As there are no triple intersections of elements of $\CU$, there are no further constraints. Together with a partition of unity $\psi_{0,1}\in U_{0,1}$, $\psi_0+\psi_1=1$ on $U_0\cap U_1$, we can define two-forms $B_0=\psi_0 F$ on $U_0$ and $B_1=-\psi_1 F$ on $U_1$ with $B_0-B_1=F$ on $U_0\cap U_1$ and $B$ is thus a curving for $A$. The corresponding curvature $H$ is globally defined and it is given by $H=\dd\psi_0\wedge F$ and $H=-\dd\psi_1\wedge F$ on $U_0$ and $U_1$, respectively. Using stokes theorem, the integral of $H$ over $S^3$ reduces to the integral of $F$ over $S^2$, confirming $H\in H^3(S^3,\RZ)$.

\subsection{Transgression and regression}

A {\em transgression map} is a map between cohomology classes on different spaces, which are the base spaces of a common correspondence space in a double fibration. We will be interested in a transgression $\CT$ from a smooth manifold $M$ to its loop space $\CL M:={\rm Map}(S^1,M)$, which actually extends from the cohomology to differential forms. The transgression map $\CT:\Omega^k(M)\rightarrow \Omega^{k-1}(\CL M)$ is based on the following double fibration:
\begin{equation}\label{dblfibrationfourself}
\begin{aligned}
\begin{picture}(50,40)
\put(0.0,0.0){\makebox(0,0)[c]{$M$}}
\put(64.0,0.0){\makebox(0,0)[c]{$\CL M$}}
\put(32.0,33.0){\makebox(0,0)[c]{$\CL M\times S^1$}}
\put(7.0,18.0){\makebox(0,0)[c]{$ev$}}
\put(62.0,18.0){\makebox(0,0)[c]{$pr$}}
\put(25.0,25.0){\vector(-1,-1){18}}
\put(37.0,25.0){\vector(1,-1){18}}
\end{picture}
\end{aligned}
\end{equation}
Here, $ev$ is the obvious evaluation map of the loop at the given angle and $pr$ desribes the projection from $\CL M\times S^1$ onto $\CL M$. The transgression map amounts to the pull back along $ev$ and the push forward along $pr$, that is $\CT=pr!\circ ev^*$, cf.\ \cite{0817647309}. Identify now the tangent space to the loop space of $M$, $T\CL M$, with the loop space of the tangent space to $M$, $\CL TM$. Note that there is a natural element $\xd(\tau)\in \CL TM$, which appears in the transgression map as follows: Given $k$ vector fields $(v_1(\tau),\ldots,v_k(\tau))$, $v_i(\tau)\in T \CL M$, and $\omega\in \Omega^{k+1}(M)$, we have
\begin{equation}
 (\CT\omega)_x(v_1(\tau),\ldots,v_k(\tau)):=\int_{S^1}\dd\tau\,\omega(v_1(\tau),\ldots,v_k(\tau),\dot{x}(\tau))~,~~~x\in\CL M~.
\end{equation}

We now embed loop space $\CL M$ into path space $\CP M$. Here, composable paths induce the notion of composable vector fields and consequently onto differential forms, cf.\ \cite{Noonan}. If we restrict ourselves to functorial forms, which are the forms respecting composability, then there is an obvious inverse map, sometimes called {\em regression}:
\begin{equation}\label{eq:regression}
 (\CT^{-1}\omega)_{x(0)}(v_0,v_1,\ldots,v_k):=\lim_{\tau\rightarrow 0}\frac{1}{\tau}\omega_{x|_{[0,\tau]}}(\tilde{v}_1,\ldots,\tilde{v}_k)~,
\end{equation}
where $v_i\in TM$ and $x\in\CP M$ such that $\xd(0)=v_0$ and $\tilde{v}_i\in\CL TM$ are extensions of the $v_i$, which are constant along the path. More details on transgressions can be found e.g.\ in \cite{0817647309} and \cite{Kiyonori:2001aa}. This map will allow us to translate the result of an ADHMN-like construction on loop space $\CL S^3$ back to $S^3$.

\section{The ADHMN construction}

In the following, we give a lightning review of the ADHMN construction \cite{Nahm:1979yw,Nahm:1981nb,Nahm:1982jb} and its D-brane interpretation \cite{Diaconescu:1996rk,Tsimpis:1998zh}. This will fix our notation and provide reference points for section 4.

\subsection{Monopoles from D1-branes ending on D3-branes}

Consider a stack of $k$ D1-branes ending on a stack of $N$ D3-branes in type IIB string theory where the worldvolumes of the branes extend into $\FR^{1,9}$ in the following way:
\begin{equation}\label{diag:D1D3}
\begin{tabular}{rcccccccc}
& 0 & 1 & 2 & 3 & 4 & 5 & 6 & \ldots\\
D1 & $\times$ & & & & & & $\times$ \\
D3 & $\times$ & $\times$ & $\times$ & $\times$ & & &
\end{tabular}
\end{equation}
Adopting the perspective of the D3-branes, the ground state of this system is effectively described by time-independent BPS configurations to $\CN=4$ super Yang-Mills theory with gauge group $\sU(N)$. The bosonic fields of this theory consist of a gauge potential $A_\mu$ and six scalar fields $\Phi^I$, $I=4,\ldots,9$. We concentrate now on configurations in the gauge $A_0=0$ for which all fields but $A_i$, $i=1,\ldots,3$, and $\Phi:=\Phi^6$ vanish. The BPS equation for these fields is given by the Bogomolny monopole equation, which is the dimensional reduction of the self-dual Yang-Mills equation $F=*F$ on $\FR^4$ (the equation describing an instanton) to $\FR^3$:
\begin{equation}\label{eq:Bogomolny}
 F=*\nabla\Phi\eor F_{ij}=\eps_{ijk}\nabla_k\Phi~,~~~i,j,k=1,\ldots,3~.
\end{equation}
Note that $\Phi$ is a harmonic function on its domain $D\subseteq\FR^3$ due to the Bianchi identity. Solutions to the Bogomolny equation with appropriate boundary conditions\footnote{amounting to the fact that the value of the Yang-Mills-Higgs action functional evaluated at the solution is finite, cf.\ \cite{Hitchin:1983ay}} carry a topological charge which is called the monopole number.

From the perspective of the D1-branes, the system is described by the Nahm equation
\begin{equation}\label{eq:Nahm}
 \dder{s}X^i=\tfrac{1}{2}\eps^{ijk}[X^j,X^k]~,
\end{equation}
where the\footnote{Throughout the paper, we will use the notation $\bar{X}:=X^\dagger$ to avoid overdecorating symbols.}  $X^i=-\bar{X}^i$ are functions on $\CI=\FR^{>0}$ with values in $\au(k)$. This equation is the dimensional reduction of the self-duality equation with gauge group $\sU(k)$ to one dimension after choosing the gauge $\nabla_s=\der{s}$.

The transition between equations \eqref{eq:Bogomolny} and \eqref{eq:Nahm}, or rather between the spaces of solutions to these two equations, is done by a certain Fourier-Mukai transform known as the Nahm transform. The latter maps self-dual gauge potentials on a four-torus to self-dual gauge potentials on the dual torus. In our special case, we are -- roughly speaking -- dealing with tori with one and three radii being infinite, respectively, and the remaining radii zero, making the two tori dual to each other. In the following, we will focus on the construction of solutions to the Bogomolny monopole equation from solutions to the Nahm equation. This construction is also known as the ADHMN construction. 

\subsection{Constructing monopoles}\label{sec:ADHMNconstruction}

We will be mostly interested in Dirac monopoles, for which the stack of D1-branes ends on a stack of D3-branes, which are all at the same position $x^6=0$. The D1-branes' worldvolume is thus $\CI=\FR^{>0}$. More commonly, one studies non-singular $\sSU(2)$-monopoles, for which the D1-branes are suspended between two D3-branes at $x^6=-1$ and $x^6=+1$ and thus $\CI=(-1,+1)$. Let $W^{n,2}(\CI)\subset L^2(\CI)$ be the Sobolev space\footnote{One often finds the notation $H^n=W^{n,2}$, which we avoid here as we label cohomology groups by $H^n$.} of functions on the interval $\CI$, which are square integrable up to their $n$th derivative. Let furthermore $X^i\in\CC^\infty(\CI,\au(k))$, $i=1,\ldots,3$, be a solution to the Nahm equation \eqref{eq:Nahm} satisfying certain boundary conditions, on which we will comment below. From these, we derive a Dirac operator $\nablas_s:W^{1,2}(\CI)\otimes\FC^2\otimes \FC^k\rightarrow W^{0,2}(\CI)\otimes\FC^2\otimes \FC^k$ together with its adjoint $\nablabs_s:W^{0,2}(\CI)\otimes\FC^2\otimes \FC^k\rightarrow (W^{1,2}(\CI)\otimes\FC^2\otimes \FC^k)^*$, which read as\footnote{We identify $W^{1,2}(\CI)\otimes\FC^2\otimes \FC^k$ with its dual $(W^{1,2}(\CI)\otimes\FC^2\otimes \FC^k)^*$.}
\begin{equation}
 \nablas_s=-\unit\dder{s}+\sigma^i\otimes \di X^i\eand \nablabs_s:=\unit\dder{s}+\sigma^i\otimes \di X^i~.
\end{equation}
Here, the $\sigma^i$ are the usual Pauli matrices with $\sigma^i=\bar{\sigma}^i$ and $X^i\in\au(k)$ with $\bar{X}^i=-X^i$. Consider now the differential operator $\Delta_s:=\nablabs_s\nablas_s$. One easily checks that the conditions $[\Delta_s,\sigma^i\otimes\unit_k]=0$, $i=1,\ldots,3$, and $\Delta_s>0$ as well as $\bar{\Delta}_s=\Delta_s$ are equivalent to the $X^i$ forming a solution to the Nahm equation. Introducing the space $\FR^3$ with euclidean coordinates $x^i$, we can shift the $\di X^i$ by $x^i\unit_k$, while preserving the properties of $\Delta_s$:
\begin{equation}
\begin{aligned}
 \nablas_{s,x}=-\unit\dder{s}+\sigma^i\otimes (\di X^i+x^i\unit_k)&\ewith \nablabs_{s,x}:=\unit\dder{s}+\sigma^i\otimes (\di X^i+x^i\unit_k)~,\\
\Delta_{s,x}:=\nablabs_{s,x}\nablas_{s,x}>0 &\eand [\Delta_{s,x},\sigma^i\otimes\unit_k]=0~. 
\end{aligned}
\end{equation}
This shift is usually called a {\em twist} of the Dirac operator, as it reflects the twisting of the original gauge bundle by the Poincar{\'e} line bundle in the underlying Nahm transform.

Consider now the orthonormalized zero modes $\psi_{s,x,\alpha}\in W^{0,2}(\CI)\otimes\FC^2\otimes \FC^k$ of the operator $\nablabs_{s,x}$:
\begin{equation}
 \nablabs_{s,x}\psi_{s,x,\alpha}=0~,~~~\alpha=1,\ldots,N~,~~~N=\dim_\FC({\rm ker}\nablabs_{s,x})~.
\end{equation}
We arrange them into a $k\times N$-dimensional matrix $\psi_{s,x}$, which satisfies
\begin{equation}
 \unit_N=\int_\CI \dd s\,\psib_{s,x}\psi_{s,x}~.
\end{equation}
Using this matrix, we define a gauge potential and a Higgs field on a subset of $\FR^3$ by
\begin{equation}\label{eq:BogomolnyFields}
A_\mu:=\int_\CI \dd s\,\psib_{s,x}\der{x^\mu}\psi_{s,x}\eand\Phi:=-\di\int_\CI \dd s\,\psib_{s,x}\,s\,\psi_{s,x}~.
\end{equation}
Note that the Green's function $G_x(s,t)$ of $\Delta_{s,x}$ with $\Delta_{s,x} G_x(s,t)=-\delta(s-t)$ is well-defined because of the positivity of $\Delta_{s,x}$. We therefore have a projector $\CP_x$ onto the orthogonal space of ${\rm ker}\nablabs_{s,x}$,
\begin{equation}\label{eq:completeness}
 \CP_x(s,t):=\nablas_{s,x} G_x(s,t)\nablabs_{t,x}=-\delta(s-t)+\psi_{s,x}\psib_{t,x}~,
\end{equation}
satisfying $\CP_x(s,t)\psi_{t,x}=0$ and $\CP^2_x(s,t)=\int \dd r\,\CP_x(s,r)\CP_x(r,t)=\CP_x(s,t)$. One can now plug the fields \eqref{eq:BogomolnyFields} into the Bogomolny equation \eqref{eq:Bogomolny}. After using \eqref{eq:completeness} to rewrite the result as double integrals over $s$ and $t$, one readily verifies that \eqref{eq:BogomolnyFields} forms a solution to the Bogomolny equation by direct computation, cf.\ e.g.\ \cite{Schenk:1986xe}. An analogous calculation is presented in detail in section \ref{ADHMNMproof}

The boundary conditions we mentioned above have to guarantee that $\dim_\FC({\ker }\nablabs_{s,x})=N$, the number of D3-branes. This is the case, if the solution $X^i$ has a simple pole at the finite boundaries of $\CI$ with its residue forming an irreducible representation of $\sSU(2)$, see e.g.\ \cite{Hitchin:1983ay} for more details.

\subsection{Abelian monopoles from the ADHMN construction}

Let us now come to the case of Dirac monopoles. The corresponding ADHMN construction is standard and found, e.g.\ in \cite{Gross:2000wc}. First, consider the simplest case $N=k=1$ and $\CI=\FR^{>0}$, i.e.\ one D1-brane ending on a single D3-brane at\footnote{One readily verifies that one can easily accommodate a translation $s\rightarrow s+s_0$ in the construction.} $x^6=s=0$. Solutions to the Nahm equation are just constants specifying the position of the monopole in the D3-brane, and we thus put $X^i=0$. The adjoint of the Dirac operator $\nablabs_{s,x}=\unit\dder{s}+x^i\sigma^i$ has the following two normalizable zero modes:
\begin{equation}
 \psi_+=\de^{-s R}\frac{\sqrt{R+x^3}}{x^1-\di x^2}\left(\begin{array}{c}{x^1-\di x^2}\\{R-x^3}\end{array} \right)\eand \psi_-=\de^{-s R}\frac{\sqrt{R-x^3}}{x^1+\di x^2}\left(\begin{array}{c}{x^3+R}\\{x^1+\di x^2}\end{array} \right)~,
\end{equation}
where $R^2=x^ix^i$. Recall that the Dirac monopole should actually be regarded as a gauge configuration on a sphere encircling the position of the monopole. The two solutions $\psi_+$ and $\psi_-$ correspond to the description of this configuration on the two standard patches of $S^2$ containing each one of the two poles at $x^3=R$ and $x^3=-R$. For $\psi_+$, we obtain the following fields from \eqref{eq:BogomolnyFields}:
\begin{equation}
 \Phi^+=-\frac{\di}{2R}\eand A^+_i=\frac{\di}{2(x^1+x^2)^2}\left(x^2\left(1-\frac{x^3}{R}\right),-x^1\left(1-\frac{x^3}{R}\right),0\right)~.
\end{equation}
They satisfy the Bogomolny equation, as one easily verifies. The zero mode $\psi_-$ yields an analogous solution which for $|x^3|\neq R$ is related to the above solution by a gauge transformation.

Next consider $k=2$, $N=1$, i.e.\ a stack of two D1-branes ending on a single D3-brane. A solution $X^i$ to the Nahm equation is found from the ansatz $X^i=f(s)T^i$ with $f\in\CC^\infty(\FR^{>0})$ and $T^i\in\asu(2)$, and we obtain
\begin{equation}\label{eq:SolutionN1k2}
 X^i=-\frac{1}{s}T^i\ewith T^i=\frac{\sigma^i}{2\di}=-\bar{T}^i~.
\end{equation}
Note that $X^i$ has indeed a simple pole at $s=0$, and the Pauli matrices $\sigma^i$ belong to an irreducible representation of $\asu(2)$. Therefore this solution can be interpreted as describing a polarization of the points of the worldvolume of the two D1-branes into fuzzy two-spheres with radius $f(s)$ and prequantum line bundle\footnote{For $k$ D1-branes, the prequantum line bundle of the fuzzy sphere is $\CO(k-1)$.} $\CO(1)$. 

The dual of the Dirac operator corresponding to the solution \eqref{eq:SolutionN1k2} reads as
\begin{equation}
 \nablabs_{s,x}=\unit\dder{s}+\sigma^i\otimes\left(-\frac{\sigma^i}{2s}+x^i\unit_2\right)~.
\end{equation}
For simplicity, let us restrict ourselves to the point $x^3=R$ and compute only the Higgs field $\Phi$:
\begin{equation}
 \psi_+=\sqrt{s}\de^{-R s}(0,0,0,1)^T~,~~~\Phi=-\frac{\di}{R}~.
\end{equation}
We get the right radial behavior and, because we started from two monopoles, the Higgs field has twice the magnitude as that of a single Dirac monopole.

\section{An ADHMN construction for self-dual strings}

\subsection{The self-dual string or M2-branes ending on M5-branes}

We now want to lift the D1-D3-brane configuration \eqref{diag:D1D3} to a configuration of a stack of M2-branes ending on M5-branes in M-theory. For this, we first have to T-dualize along one direction transverse to both D1- and D3-branes, yielding a D2-D4-brane configuration. Subsequently, we choose an M-theory direction along which we can wrap the M5-brane corresponding to the D4-brane:
\begin{equation}\label{diag:Branes}
\begin{aligned}
\begin{tabular}{rccccccc}
${\rm M}$ & 0 & 1 & 2 & 3 & \phantom{(}4\phantom{)} & 5 & 6 \\
M2 & $\times$ & & & & & $\times$ & $\times$ \\
M5 & $\times$ & $\times$ & $\times$ & $\times$ & $\times$ & $\times$ &
\end{tabular}&\\[0.2cm]
\downarrow S^1_{\rm M}\hspace{1.15cm}&\\[0.2cm]
\begin{tabular}{rccccccc}
IIA & 0 & 1 & 2 & 3 & (4) & 5 & 6 \\
D2 & $\times$ & & & & & $\times$ & $\times$ \\
D4 & $\times$ & $\times$ & $\times$ & $\times$ & & $\times$ &
\end{tabular}&
~\stackrel{T_5}{\longleftrightarrow}~
\begin{tabular}{rccccccc}
IIB & 0 & 1 & 2 & 3 & (4) & 5 & 6 \\
D1 & $\times$ & & & & & & $\times$ \\
D3 & $\times$ & $\times$ & $\times$ & $\times$ & & &
\end{tabular}
\end{aligned}
\end{equation}
Note that here, we embedded the target space of string theory, $\FR^{1,9}$, into that of M-theory, $\FR^{1,10}$, as the hyperplane $x^4=0$. In the following, we will restrict our discussion to the cases for which we have reasonable worldvolume theories, i.e.\ to the cases of a single M2-brane ending on a single M5-brane, $N=k=1$, and the case of two M2-branes ending on a single M5-brane, $N=1$, $k=2$.

The equations of motion of a single M5-brane can be derived in a number of ways, e.g.\ by the doubly supersymmetric approach to super $p$-branes or by analyzing the dynamics of the Goldstone modes arising from the symmetry breaking of the 11d supergravity action by the presence of the M5-brane, see e.g.\ \cite{Berman:2007bv} for a review. From analyzing the Goldstone modes, we learn that the fields on the M5-brane are given by five scalars $\Phi^I$ together with a self-dual two-form potential $B$, such that $H:=d B=*H$. The doubly supersymmetric approach was pursued for example in \cite{Howe:1997ue}, where also the appropriate BPS equation corresponding to a stack of M2-branes ending on a single M5-brane as in \eqref{diag:Branes} was given:
\begin{equation}\label{eq:SelfDualString}
 H_{05\mu}=\tfrac{1}{4}\dpar_\mu\Phi,~~~H_{\mu\nu\rho}=\tfrac{1}{4}\eps_{\mu\nu\rho\sigma}\dpar_\sigma\Phi~,~~~\mu,\nu,\rho,\sigma=1,\ldots,4~,
\end{equation}
Here, $\Phi=\Phi^6$ is a function which is harmonic on its domain $D\subseteq\FR^4$ due to the Bianchi identity. We therefore have $\Phi=\Phi_0+\sum_p\frac{1}{|x-y_p|^2}$ with $D=\FR^4\backslash \{y_p\}$, where $y_p$ are the singular points of $\Phi$ corresponding to positions of M2-brane boundaries. Because of the self-duality of $H$ and the string-like shape of the na{\"i}ve boundary of the M2-brane on the M5-brane, this configuration is known as the self-dual string soliton. Note that $\Phi\sim\frac{1}{R^2}$, contrary to the case of the D1-D3-brane system, where we had $\Phi\sim\frac{1}{R}$.

From the perspective of the M2-branes, an effective description was not available until Basu and Harvey proposed an extension of the Nahm equation \cite{Basu:2004ed}. They suggested that the field content is given by four transverse scalars $X^\mu$, which take values in a 3-Lie algebra\footnote{The definition of a metric 3-Lie algebra is found in appendix A.} $\CA$. Basu and Harvey then postulated essentially the following extension of the Nahm equation \eqref{eq:Nahm}:
\begin{equation}\label{eq:BasuHarvey}
 \dder{s}X^\mu=\tfrac{1}{3!}\eps^{\mu\nu\rho\sigma}[X^\nu,X^\rho,X^\sigma]~,~~~X^\mu\in\CA~,
\end{equation}
which can be justified by $\sSO(4)$-invariance and dimensional analysis: Recall that $s$ and $X^\mu$ correspond to $\Phi\sim \frac{1}{r^2}$ and $x^\mu$, respectively, in the self-dual string equation \eqref{eq:SelfDualString}. 

\subsection{The Basu-Harvey equation from a twisted Dirac operator}

Let us now motivate a Dirac operator suitable for an extended ADHMN construction for the M2-M5-brane system by following the Dirac operator of the D1-D3-brane system through T-duality and the lift to M-theory.

Our starting point is the Dirac operator $\nablas^{\rm IIB}_{s,x}=-\unit\dder{s}+\sigma^i\otimes (\di X^i+x^i\unit_k)$ in type IIB string theory. Going from a chiral theory to a non-chiral one (type IIA), it is only natural to replace the Pauli matrices with the generators $\gamma^\mu$ of the Clifford algebra\footnote{Our conventions are $\{\gamma^\mu,\gamma^\nu\}=+2\delta^{\mu\nu}$.} $Cl(\FR^4)$. Such a Dirac operator was suggested e.g.\ in \cite{Campos:2000de}, and here we choose:
\begin{equation}\label{eq:DiracOperatorIIA}
 \nablas_{s,x}^{\rm IIA}=-\gamma_5\otimes\unit_k\dder{s}+\gamma^4\gamma^i\otimes (X^i-\di x^i)~,~~~\nablabs_{s,x}^{\rm IIA}=\gamma_5\otimes\unit_k\dder{s}+\gamma^4\gamma^i\otimes (X^i-\di x^i)~,
\end{equation}
where $\gamma_5:=\gamma^1\gamma^2\gamma^3\gamma^4$. One readily verifies that $\Delta^{\rm IIA}_{s,x}:=\nablabs_{s,x}^{\rm IIA}\nablas^{\rm IIA}_{s,x}>0$ and $[\Delta^{\rm IIA}_{s,x},\gamma^4\gamma^i]=0$ amounts again to the $X^i$ satisfying the Nahm equation \eqref{eq:Nahm}. Note, however, that we have $\dim({\rm ker} \nablabs_{s,x}^{\rm IIA})=2\dim({\rm ker}\nablabs_{s,x}^{\rm IIB})$, as the Dirac operator acts reducibly:
\begin{equation}
 \nablas_{s,x}^{\rm IIA}=\left(\begin{array}{cc}
\nablas_{s,x}^{\rm IIB} & 0 \\
0 & -\nablas_{s,x}^{\rm IIB}
\end{array}
\right)~.
\end{equation}

To lift \eqref{eq:DiracOperatorIIA} to M-theory, recall that solutions $X^\mu$ to the Basu-Harvey equation \eqref{eq:BasuHarvey}, which will play the r{\^o}le of our Nahm data, take values in a 3-Lie algebra $\CA$. We therefore expect the M-theory Dirac operator $\nablas_s^{\rm M}$ to act on elements of a space $\FC^4\otimes \CE \otimes W^{1,2}(\FR^{>0})$, where $\CE$ is a space related to the 3-Lie algebra $\CA$. There are now essentially two possibilities for the choice of $\CE$: an analogue of the carrier space of the adjoint representation, i.e.\ $\CE=\CA$, and an analogue of the fundamental representation, i.e.\ $\CE=\FC^d$. Both possibilities, when followed through, yield construction mechanisms for self-dual strings. However, a future extension to the non-abelian case seems to work only with the latter possibility, and we will therefore choose $\CE=\FC^d$, following closely the original ADHMN construction.

The 3-Lie algebra $\CA$ comes with an associated Lie algebra of inner derivations, which we denote by $\frg_{\CA}$. Let $\rho$ be a faithful representation of $\frg_{\CA}$ that is compatible with the 3-Lie algebra structure as described in appendix A and let $\FC^d$ be its carrier space. For example, $\frg_{A_4}$ admits the following compatible representation with $d=4$:
\begin{equation}
 D^{(\rho)}(e_\alpha, e_\beta)\zeta=\tfrac{1}{2}\gamma^{\alpha\beta}\gamma_5 \zeta~,~~~\zeta \in \FC^4~,
\end{equation}
where the $e_\alpha$, $\alpha=1,\ldots,4$, form an orthonormal basis of the 3-Lie algebra $A_4$. Dimensional analysis as well as $\sSO(4)$-invariance then naturally lead us to the (untwisted) Dirac operator
\begin{equation}
 \nablas^{\rm M}_s=-\gamma_5\dder{s}+\tfrac{1}{2}\gamma^{\mu\nu}D^{(\rho)}(X^\mu,X^\nu)\ewith\nablabs^{\rm M}_s=\gamma_5\dder{s}+\tfrac{1}{2}\gamma^{\mu\nu}D^{(\rho)}(X^\mu, X^\nu)~.
\end{equation}
Here, $\gamma^{\mu\nu}:=\tfrac{1}{2}[\gamma^\mu,\gamma^\nu]$, $X^\mu\in\CA$ and $D^{(\rho)}(X^\mu, X^\nu)$ is the inner derivation $D(X^\mu,X^\nu)\in\frg_\CA$ in the representation $\rho$. The Dirac operator $\nablas^{\rm M}_s$ thus acts on elements of $\FC^4\otimes\FC^d\otimes W^{1,2}(\FR^{>0})$. The differential operator $\Delta^{\rm M}_s:=\nablabs^{\rm M}_s\nablas^{\rm M}_s$ reads explicitly as
\begin{equation}
\Delta^{\rm M}_s=-\unit\left(\dder{s}\right)^2+\tfrac{1}{2}\gamma_5\gamma^{\mu\nu}\dder{s}D^{(\rho)}(X^\mu,X^\nu)+\tfrac{1}{2^2}\gamma^{\mu\nu}\gamma^{\kappa\lambda}D^{(\rho)}(X^\mu,X^\nu) D^{(\rho)}(X^\kappa,X^\lambda)~,
\end{equation}
and a straightforward calculation shows that the condition $[\Delta^{\rm M}_s,\gamma^{\mu\nu}]=0$ for all $\mu,\nu=1,\ldots,4$ amounts to the Basu-Harvey equation \eqref{eq:BasuHarvey}. 

The key issue in extending the ADHMN construction to self-dual strings is the appropriate twist of the Dirac operator. It is clear that twisting has to amount to adding a term $\gamma^{\mu\nu}a^\mu b^\nu$ to $\nablas_s^{\rm M}$, where the vectors $a$ and $b$ must have the same dimension as $x$. 

Recall that the two-form potential $B$, which we want to construct and which describes together with the Higgs field $\Phi$ the self-dual string, actually belongs to the connective structure of a gerbe. Using a transgression map, we can switch from the gerbe to a $\sU(1)$-bundle over loop space and perform our construction there. A regression map can take us back to the gerbe picture afterwards. Similar to a Dirac monopole corresponding to a vector bundle over $S^2$ instead of $\FR^3$, we expect here a gerbe over $S^3$ instead of $\FR^4$. The transgression map therefore takes us to loops on $S^3$, which we describe as embedded in $\FR^4$ by the cartesian coordinates $x^\mu$. We thus have loops $x^\mu(\tau)$ satisfying
$x^\mu(\tau)x^\mu(\tau)=R^2$ and $x^\mu(\tau)\xd^\mu(\tau)=0$, where $R$ is the radius of $S^3\subset \FR^4$ and $\xd^\mu(\tau):=\dderr{x^\mu(\tau)}{\tau}$. We also impose the following (gauge) condition on the parameterization of our loops: $\xd^\mu(\tau)\xd^\mu(\tau)=R^2$. Now there is an obvious twist of the Dirac operator:
\begin{equation}
\begin{aligned}
 \nablas^{\rm M}_{s,x(\tau)}&=-\gamma_5\dder{s}+\gamma^{\mu\nu}\left(\tfrac{1}{2}D^{(\rho)}(X^\mu,X^\nu)-\di x^\mu(\tau)\xd^\nu(\tau)\right)~.
\end{aligned}
\end{equation}
Note that the Nahm data $(X^\mu)$ is {\em not} extended to Nahm data on the circle parameterized by $\tau$. Moreover, note that the twist naturally reflects the fact that $x(\tau)\in\CL S^3$: The antisymmetrization of $x^{\mu}(\tau)\xd^{\nu}(\tau)$ eliminates a possible component of $\xd^{\nu}(\tau)$ parallel to $x^\mu(\tau)$. The condition that $\Delta^{\rm M}_{s,x(\tau)}:=\nablabs^{\rm M}_{s,x(\tau)}\nablas^{\rm M}_{s,x(\tau)}$ commutes with all the $\gamma^{\mu\nu}$ is again equivalent to the $X^\mu$ satisfying the Basu-Harvey equation \eqref{eq:BasuHarvey}. 

\subsection{The self-dual string on loop space}

Now that we have a Dirac operator connected to loop space, we need to use a transgression map to translate the self-dual string equation $H=*\dd \Phi$ to loop space, as well. Although there is no Hodge star operation due to the loop space having infinite dimension, one can use the transgression map and its inverse to lift the Hodge star on $\FR^4$ to an operation on $\CL \FR^4$. The self-dual string equation in \eqref{eq:SelfDualString}, as an equation on three-forms, reads as:
\begin{equation}\label{eq:sdsexplicit}
 H=\left(\eps_{\mu\nu\rho\sigma}\der{x^\sigma}\Phi\right)\dd x^\mu\wedge \dd x^\nu\wedge \dd x^\rho~.
\end{equation}
The transgression map $\CT$ now maps $H\in\Omega^3(\FR^4)$ to $F:=\CT(H)\in\Omega^2(\CL\FR^4)$ according to 
\begin{equation}
\begin{aligned}
F(V_1,V_2)=&\int_{S^1} \dd \tau\, F_{\mu\nu}(x(\tau))\;V^\mu_1(x(\tau))\;V^\nu_2(x(\tau))\\
:=&\int_{S^1} \dd\tau\, H_{\mu\nu\rho}(x(\tau))\;V^\mu_1(x(\tau))\;V^\nu_2(x(\tau))\;\xd^\rho(\tau)~,
\end{aligned}
\end{equation}
where $V_1, V_2$ are elements of $T\CL\FR^4$. Equation \eqref{eq:sdsexplicit} translates correspondingly into
\begin{equation}
\begin{aligned}
 \int_{S^1} \dd\tau\, H_{\mu\nu\rho}(x(\tau))\;V^\mu_1(x(\tau))\;V^\nu_2(x(\tau))\;\xd^\rho(\tau)&=\\\int_{S^1} \dd\tau\, \eps_{\mu\nu\rho\sigma}\;V^\mu_1(x(\tau))\;V^\nu_2(x(\tau))\;&\xd^\rho(\tau)\int \dd \varrho\,\derdel{x^\sigma(\varrho)}\;\Phi(x(\tau))~,
\end{aligned}
\end{equation}
where $\derdel{x^\nu(\varrho)}x^\mu(\sigma)=\delta^\mu_\nu\delta(\varrho-\sigma)$. The transgression map $\CT$ also takes the two-form potential $B\in\Omega^2(\FR^4)$ to a gauge potential $A\in\Omega^1(\CL \FR^4)$, which acts onto $V\in T\CL\FR^4$ according to
\begin{equation}
 A(V)=\int_{S^1} \dd \tau\, A_\mu(x(\tau))\;V^\mu(x(\tau)):=\int_{S^1} \dd\tau\, B_{\mu\nu}\;V^\mu(x(\tau))\;\xd^\nu(\tau)~.
\end{equation}
Note that since $\dpar S^1=\emptyset$, transgression is in our case a chain map:
\begin{equation}
 F:=\delta A=\delta \CT (B)=\CT (\dd B)=\CT(H)~,
\end{equation}
where $\delta f:=\int \dd\tau \frac{\delta f}{\delta x^\mu(\tau)} \delta x^\mu(\tau)$ is the loop space\footnote{Differential calculus on loop spaces can be made precise within the context of diffeological spaces or as done by K.-T. Chen, cf.\ e.g.\ the references in \cite{Stacey:2008aa}.} differential. Here, we will consider a different gauge potential $A$ such that $F_{\mu\nu}=\dpar_\mu A_\nu-\dpar_\nu A_\mu$, where $\dpar_\mu= \int \dd\rho \frac{\delta}{\delta x^\mu(\rho)}$. This guarantees that we obtain a field strength $F$ corresponding to a transgressed 3-form $H$.

Putting everything together, we arrive at the following equation on loop space:
\begin{equation}\label{eq:LoopSpaceSDS}
 F_{\mu\nu}(x(\tau)):=\der{x^\mu}A_\nu(x(\tau))-\der{x^\nu}A_\mu(x(\tau))=\eps_{\mu\nu\rho\sigma}\xd^\rho(\tau) \der{x^\sigma}\Phi(x(\tau))~.
\end{equation}
We call \eqref{eq:LoopSpaceSDS} the loop space self-dual string equation and we will find solutions to this equation via a generalization of the ADHMN construction involving the Dirac operator we constructed above. 

In the following, we suppress the integral over $S^1$, as we have done in \eqref{eq:LoopSpaceSDS}; all our equations already hold in this form.

\subsection{Constructing self-dual strings}\label{ADHMNMproof}

Our construction proceeds now in the obvious way, cf.\ section \ref{sec:ADHMNconstruction} Consider the Dirac operator defined above, where we use again the compatible representation $\rho$ of $\frg_\CA$ with $d$-dimensional carrier space:
\begin{equation}
 \nablas_{s,x(\tau)}^{\rm M}:\FC^4\otimes\FC^d\otimes W^{1,2}(\CI)\rightarrow \FC^4\otimes \FC^d\otimes W^{0,2}(\CI)~,~~~\CI=(0,\infty)~.
\end{equation}
We pick one of the normalizable zero modes of $\nablabs_{s,x(\tau)}^{\rm M}$, denote it by $\psi_{s,x(\tau)}$,
\begin{equation}\label{eq:normalizePsiM}
 1=\int_\CI\dd s\, \psib_{s,x(\tau)}\psi_{s,x(\tau)}~,
\end{equation}
and introduce the following fields on loop space:
\begin{equation}\label{eq:MFieldDefinitions}
 A_\mu(x(\tau))=\int \dd s\, \psib_{s,x(\tau)} \der{x^\mu} \psi_{s,x(\tau)}\eand\Phi(x(\tau))=-\di\int \dd s\, \psib_{s,x(\tau)}\,s\,\psi_{s,x(\tau)}~.
\end{equation}
We will now show that these fields solve indeed the loop space self-dual string equation \eqref{eq:LoopSpaceSDS} if the $X^\mu$ which determine the Dirac operator $\nablas^{\rm M}_{s,x(\tau)}$ satisfy the Basu-Harvey equation. Recall that in this situation, the differential operator $\Delta^{\rm M}_{s,x(\tau)}:=\nablabs_{s,x(\tau)}^{\rm M}\nablas_{s,x(\tau)}^{\rm M}$ is a linear combination of $\unit$ and $\gamma_5$ (i.e.\ it commutes with any $\gamma^{\mu\nu}$). In order to have a Green's function $G^{\rm M}_{x(\tau)}(s,t)$ with $\Delta^{\rm M}_{s,x(\tau)} G^{\rm M}_{x(\tau)}(s,t)=-\delta(s-t)$, we also require this differential operator to be invertible. To verify this, consider an arbitrary $\psi\in\FC^4\otimes\FC^d\otimes W^{1,2}(\CI)$ with $(\psi,\psi)>0$, where $(\cdot,\cdot)$ is the obvious combination of hermitian scalar product on the complex vector spaces and the $L^2$-norm over $W^{1,2}(\CI)$. Taking into account that the Nahm data satisfies the Basu-Harvey equation, it remains to show that
\begin{equation}\label{eq:invertibility}
 \left(\psi,\left(-\left(\dder{s}\right)^2+\tfrac{1}{2}\{\gamma^{\mu\nu},\gamma^{\rho\sigma}\}\otimes T^{\mu\nu} T^{\rho \sigma}\right)\psi\right)>0~,
\end{equation}
where $T^{\mu\nu}:=\left(\tfrac{1}{2}D^{(\rho)}(X^\mu,X^\nu)-\di x^{[\mu}(\tau)\xd^{\nu]}(\tau)\right)$. Note that $\bar{T}^{\mu\nu}=-T^{\mu\nu}$ and that there is no $\psi$ such that\footnote{The representation $\rho$ is faithful and $x(\tau)\in\CL S^3$.} $T^{\mu\nu}\psi=0$ for all $\mu,\nu$. We can simplify \eqref{eq:invertibility} further to
\begin{equation}\label{eq:resultinequality}
\begin{aligned}
 2\left(T^{\mu\nu}\psi,T^{\mu\nu}\psi\right)+2\left(\tfrac{1}{2}\eps^{\mu\nu\rho\sigma}T^{\rho\sigma}\psi,\tfrac{1}{2}\eps^{\mu\nu\kappa\lambda}T^{\kappa\lambda}\psi\right)
 -4\left(T^{\mu\nu}\psi,\gamma_5\tfrac{1}{2}\eps^{\mu\nu\rho\sigma}T^{\rho\sigma}\psi\right)>0~.
\end{aligned}
\end{equation}
Next, we decompose $\psi$ into eigenvectors $\psi_\pm$ of $\gamma_5\otimes \unit_d$ with eigenvalues $\pm 1$. These eigenspaces are obviously orthogonal. Because of $[\gamma_5\otimes \unit_d,T^{\mu\nu}]=0$, $T^{\mu\nu}\psi_\pm$ belongs to the same eigenspace as $\psi_\pm$. Relation \eqref{eq:resultinequality} now reduces to
\begin{equation}
\begin{aligned}
 \left(T^{\mu\nu}\psi_+-\tfrac{1}{2}\eps^{\mu\nu\rho\sigma}T^{\rho\sigma}\psi_+~,~T^{\mu\nu}\psi_+-\tfrac{1}{2}\eps^{\mu\nu\kappa\lambda}T^{\kappa\lambda}\psi_+\right)&\\+\left(T^{\mu\nu}\psi_-+\tfrac{1}{2}\eps^{\mu\nu\rho\sigma}T^{\rho\sigma}\psi_-~,~T^{\mu\nu}\psi_-+\tfrac{1}{2}\eps^{\mu\nu\kappa\lambda}T^{\kappa\lambda}\psi_-\right)&>0~,
\end{aligned}
\end{equation}
which holds true if $(\psi,\psi)>0$, because the $T^{\mu\nu}$ do not have a common kernel. Therefore $\Delta_{s,x(\tau)}^{\rm M}$ is invertible and has a Green's function, which leads again to a projector, cf.\ \eqref{eq:completeness}:
\begin{equation}
 \CP^{\rm M}_{x(\tau)}(s,t):=\nablas^{\rm M}_{s,x(\tau)} G^{\rm M}_{x(\tau)}(s,t)\nablabs^{\rm M}_{t,x(\tau)}=-\delta(s-t)+\psi_{s,x(\tau)}\psib_{t,x(\tau)}~,
\end{equation}
where we switched to the notation $(\psi,A \psi)=\psib A \psi$ for convenience. For the same reason, let us also temporarily drop the explicit $x(\tau)$-dependence and write $x$ for $x(\tau)$ as well as $\dpar_\mu$ for $\int \dd \varrho \derdel{x^\mu(\varrho)}$. We have
\begin{equation}
 \begin{aligned}
  F_{\mu\nu}&=\int \dd s\,(\dpar_{[\mu}\psib_s)\dpar_{\nu]}\psi_s\\
&=\int \dd s\int \dd t\,(\dpar_{[\mu}\psib_s)\left(\psi_s\psib_t-\nablas_s^{\rm M}G^{\rm M}(s,t)\nablabs_t^{\rm M}\right)\dpar_{\nu]}\psi_t\\
&=\int \dd s\int \dd t\,\psib_s\left(\gamma^{\mu\kappa}\xd^\kappa G^{\rm M}(s,t)\gamma^{\nu\lambda}\xd^\lambda-\gamma^{\nu\kappa}\xd^\kappa G^{\rm M}(s,t)\gamma^{\mu\lambda}\xd^\lambda\right)\psi_t~.
 \end{aligned}
\end{equation}
Recall that the Green's function commutes with the $\gamma^{\mu\nu}$ and $\gamma_5$. Together with the identity
\begin{equation}
 [\gamma^{\mu\kappa},\gamma^{\nu\lambda}]\xd^\kappa\xd^\lambda=-2\eps_{\mu\nu\rho\sigma}\gamma^{\sigma\kappa}\gamma_5\xd^\rho\xd^\kappa~,
\end{equation}
we thus arrive at
\begin{equation}
 \begin{aligned}
F_{\mu\nu}&=-\eps_{\mu\nu\rho\sigma}\int \dd s\int \dd t\,\psib_s\left(2\gamma^{\sigma\kappa}\gamma_5G^{\rm M}(s,t)\xd^\rho\xd^\kappa\right)\psi_t\\
&=-\di\eps_{\mu\nu\rho\sigma}\xd^\rho\int \dd s\int \dd t\,\Big((\dpar_\sigma \psib_s)\left(\psi_s\psib_t-\nablas^{\rm M}_sG^{\rm M}(s,t)\nablabs^{\rm M}_t\right)\,t\,\psi_t+\\&\hspace{4.5cm}\psib_s\,s\,\left(\psi_s\psib_t-\nablas^{\rm M}_sG^{\rm M}(s,t)\nablabs^{\rm M}_t\right)\dpar_\sigma \psi_t\Big)\\
&=-\di\eps_{\mu\nu\rho\sigma}\xd^\rho\int \dd s\,(\dpar_\sigma\psib_s)\,s\,\psi_s+\psib_s\,s\,\dpar_\sigma\psi_s\\
&=\eps_{\mu\nu\rho\sigma}\xd^\rho\dpar_\sigma\Phi~,
 \end{aligned}
\end{equation}
thus verifying that the fields \eqref{eq:MFieldDefinitions} indeed solve the loop space self-dual string equation \eqref{eq:LoopSpaceSDS}.

\subsection{Explicit solutions}

Let us now come to explicit solutions. We restrict ourselves again to the cases $N=k=1$ and $N=1$, $k=2$, as we did for monopoles. For $k=1$, the Nahm data reduces to constants and we therefore put $X^\mu=0$. The equation $\nablabs^{\rm M}_{s,x(\tau)}\psi_{s,x(\tau)}=0$ and the normalization condition \eqref{eq:normalizePsiM} are enough to fix $\psi_{s,x(\tau)}$ completely. One finds altogether eight zero modes. Half of them are normalizable on $(0,\infty)$, the other half is normalizable on $(-\infty,0)$. The remaining four split into two pairs, each giving the description on one of the two standard patches of $S^3$. Recalling that there was a doubling of the zero modes due to switching from Pauli matrices to generators of the Clifford algebra $Cl(\FR^4)$, we restrict ourselves to the zero mode $\psi_{s,x(\tau)}$ with $\gamma_5\psi_{s,x(\tau)}=\psi_{s,x(\tau)}$. Up to normalization, we have\footnote{For brevity, we write $x^\mu$ for $x^\mu(\tau)$ and $\xd^\mu$ for $\xd^\mu(\tau)$.}:
\begin{equation}
\psi_{s,x(\tau)}\sim\de^{-R^2 s}\left(\begin{array}{cc}
\di \left(R^2+x^2 \xd^1-x^1 \xd^2-x^4 \xd^3+x^3 \xd^4\right) \\
x^3 (\xd^1+\di \xd^2)+x^4 (\xd^2-\di \xd^1)-(x^1+\di x^2) (\xd^3-\di \xd^4) \\
0\\ 
0                                
\end{array}
\right)~,
\end{equation}
and formulas \eqref{eq:MFieldDefinitions} yield for the normalized zero mode:
\begin{equation}\label{eq:4.28}
\begin{aligned}
 &\hspace{5cm}\Phi(x)=\frac{\di}{2 R^2}~,\\[0.2cm]
 A(x)&=\frac{1}{n(x)}\left(\begin{array}{c}
-x^3 (\xd^2 \xd^3 + \xd^1 \xd^4) + x^4 (\xd^1 \xd^3 - \xd^2 \xd^4) +
  x^2 ((\xd^3)^2 + (\xd^4)^2)\\ 
x^4 (\xd^2 \xd^3 + \xd^1 \xd^4) + x^3 (\xd^1 \xd^3 - \xd^2 \xd^4) - 
 x^1 ((\xd^3)^2 + (\xd^4)^2)\\ -R^2 x^4 + 
 \xd^3 (x^1 \xd^2 -x^2 \xd^1 + x^4 \xd^3 - x^3 \xd^4)\\
R^2 x^3 + \xd^4 (x^1 \xd^2 -x^2 \xd^1 + x^4 \xd^3 - x^3 \xd^4)          
\end{array}
\right)~,
\end{aligned}
\end{equation}
where $n(x)=-2\di R^2(R^2-x^2\xd^1+x^1\xd^2+x^4\xd^3-x^3\xd^4)$. These fields indeed solve the self-dual string equation on loop space \eqref{eq:LoopSpaceSDS}. Let us now switch to spherical coordinates $(\theta^1,\theta^2,\phi)$ describing $S^3\subset \FR^4$ according to
\begin{equation*}
 x^1=R\sin\theta^1\sin\theta^2\cos\phi~,~~x^2=R\sin\theta^1\sin\theta^2\sin\phi~,~~x^3=R\sin\theta^2\cos\theta^1~,~~x^4=R\cos\theta^2~.
\end{equation*}
In these coordinates, the field strength $F$ of the above gauge potential $A$ reads as
\begin{equation}\label{eq:GaugeFieldLoopSpace}
 F=\frac{2 \di \sin\theta^1 \sin^2\theta^2 (\dot{\theta}^2\,\dd\phi\wedge\dd\theta^1 - \dot{\theta}^1 \,\dd\phi\wedge\dd\theta^2 + 
   \dot{\phi} \,\dd\theta^1\wedge\dd\theta^2)}{\sqrt{\dot{\phi}^2 + 2 (\dot{\theta}^1)^2 + 
 4 (\dot{\theta}^2)^2 - (\dot{\phi}^2 + 2 (\dot{\theta}^1)^2) \cos(2\theta^2) - 2 \dot{\phi}^2 \cos(2\theta^1) \sin^2\theta^2}}~,
\end{equation}
and the induced metric on $S^3$ is given by
\begin{equation}
 \dd s^2=R^2\sin^2\theta^2\,\dd\theta^1\otimes\dd\theta^1+R^2\,\dd\theta^2\otimes\dd\theta^2+R^2\sin^2\theta^1\sin^2\theta^2\,\dd\phi\otimes\dd\phi~.
\end{equation}
We can now compute the regression $H=\CT^{-1}F$ of \eqref{eq:GaugeFieldLoopSpace} back to $S^3$ using formula \eqref{eq:regression}:
\begin{equation}
\begin{aligned}
 H=&F|_{\dot{\theta}^1=1,\dot{\theta}^2=0,\dot{\phi}=0}\wedge \sin\theta^2\dd\theta^1\\&~-F|_{\dot{\theta}^1=0,\dot{\theta}^2=1,\dot{\phi}=0}\wedge \dd\theta^2\\&~+F|_{\dot{\theta}^1=0,\dot{\theta}^2=0,\dot{\phi}=1}\wedge \sin\theta^1\sin\theta^2\dd\phi\\
=&6\di\sin\theta^1\sin^2\theta^2\,\dd \theta^1\wedge \dd\theta^2\wedge \dd\phi~,
\end{aligned}
\end{equation}
and one recovers\footnote{To understand the above regression, it might help to perform the inverse operation and to compute the transgression $F=\CT(H)$. Recall the normalization condition $|\dot{x}|^2=R^2$ we imposed on the parameterization of the loops.}
the 3-form field strength of the self-dual string on $S^3\subset \FR^4$. 

For the case $k=2$, we start from the following solution to the Basu-Harvey equation:
\begin{equation}
 X^\mu=\frac{e_\mu}{\sqrt{2s}}~,
\end{equation}
where the $e_\mu$ are the orthonormalized generators of $A_4$. This solution can be interpreted as each point of the worldvolume of the M2-branes polarizing into a fuzzy 3-sphere with radius $r\sim\frac{1}{\sqrt{2s}}$. Plugging this solution into the Dirac operator, we can again calculate the corresponding zero modes. As the computation is rather cumbersome in practice, we restrict ourselves to the value of the Higgs field $\Phi$ at $x^1(\tau)=\xd^2(\tau)=R$, which we then extend by $\sSO(4)$-invariance. The corresponding zero mode reads as
\begin{equation}
 \psi_+(x)=4 R^2\sqrt{s}\de^{-2R^2 s}(1,0,0,0,0,0,0,0,0,0,0,0,0,0,0,0)^T~,
\end{equation}
and formulas \eqref{eq:MFieldDefinitions} yield a Higgs field with the right behavior,
\begin{equation}
 \Phi(x)=\frac{\di}{R^2}~.
\end{equation}

\subsection{Reduction to the Nahm equation}

To anchor our construction deeper within the context of the BLG model and the ADHMN construction, let us describe how it reduces to the corresponding construction for the D2-D4-brane system. For this, recall the reduction from the BLG model with 3-Lie algebra $A_4$ to $\CN=8$ super Yang-Mills theory in three dimensions with gauge group $\sSU(2)$ as developed in \cite{Mukhi:2008ux}. The key here was to give the scalar corresponding to the M-theory direction $x^\natural$ a vacuum expectation value: 
\begin{equation}\label{eq:MukhiReduction}
 \langle X^\natural\rangle =\frac{r}{\ell_p^{3/2}}e_4=g_{\rm YM}e_4~,
\end{equation}
where $r$ is the radius of the circle on which $x^\natural$ is compactified, $\ell_p$ is the Planck length, $g_{\rm YM}$ the Yang-Mills coupling and $e_4$ a chosen generator of $A_4$. After a duality transform of the gauge potential, the terms to leading order in $g_{\rm YM}$ yield the action of three-dimensional maximally supersymmetric Yang-Mills theory.

In our case, we start by demanding that $X^\natural=X^4=-Re_4$, in close analogy with \eqref{eq:MukhiReduction}. We then pick a loop $x(\tau)$ and a base point $x(\tau_0)$ such that $\xd^4(\tau_0)=-R$. From our normalization of $\xd(\tau)$, it follows that $\xd^i(\tau_0)=0$ for $i=1,\ldots,3$ and from the orthogonality $\xd^\mu(\tau)x^\mu(\tau)=0$, we conclude that $x^4(\tau_0)=0$. That is, at $\tau_0$ the loop agrees with a circle around the M-theory direction. We consider now the Dirac operator $\nablas^{\rm M}_{s,x(\tau)}$ at this loop with base point, thereby going from $\CL S^3$ back to the correspondence space $\CL S^3\times S^1$ in the double fibration \eqref{dblfibrationfourself} with $M=S^3$. To reduce it further to $S^3$, we have to evaluate it at the base point of the loop:
\begin{equation}\label{eq:DiracReduction}
\begin{aligned}
 \nablas^{\rm M}_{s,x(\tau)}=-&\gamma_5\dder{s}+\gamma^{\mu\nu}\left(\tfrac{1}{2}D^{(\rho)}(X^\mu,X^\nu)-\di x^\mu(\tau)\xd^\nu(\tau)\right)\\~\rightarrow~&-\gamma_5\dder{s}+\gamma^{\mu\nu}\left(\tfrac{1}{2}D^{(\rho)}(X^\mu,X^\nu)-\di x^\mu(\tau_0)\xd^\nu(\tau_0)\right)\\~~~=&-\gamma_5\dder{s}+R\gamma^{4 i}\left(X^{i \alpha}D^{(\rho)}(e_\alpha,e_4)-\di x^i(\tau_0)\right)~.
\end{aligned}
\end{equation}
Note that $D(e_4,e_4)=0$ and the inner derivations $D(e_\alpha,e_4)$ in the representation $\rho$ of $\frg_{A_4}$ with carrier space $\FC^4$ given in appendix A form a (reducible) representation of $\asu(2)$:
\begin{equation}
 [D^{(\rho)}(e_\alpha,e_4),D^{(\rho)}(e_\beta,e_4)]=\eps_{\alpha\beta\gamma}D^{(\rho)}(e_\gamma,e_4)~,~~~\alpha,\beta,\gamma=1,\ldots,3~.
\end{equation}
Also recall that $s^{-1/2}\sim R$ for the self-dual string but $s^{-1}\sim R$ for the monopole. We thus recognize the Dirac operator $\nablas^{\rm IIA}_{s,x}$ in \eqref{eq:DiracReduction}. On the level of equations of motion, the Basu-Harvey equation reduces to the Nahm equation according to
\begin{equation}
 \begin{aligned}
    \dder{s}X^\mu&=\tfrac{1}{3!}\eps^{\mu\nu\rho\sigma}[X^\nu,X^\rho,X^\sigma]~~\rightarrow~~\dder{s}X^i=\tfrac{1}{2}\eps^{ijk}R[X^j,X^k]~.
 \end{aligned}
\end{equation}
Eventually, note that $A_\mu(x(\tau))=B_{\mu\nu}\xd^\nu\rightarrow R A_i(\tau_0)$ and therefore the loop space self-dual string equation reduces as follows:
\begin{equation}
 F_{\mu\nu}(x(\tau))=\eps_{\mu\nu\rho\sigma}\xd^\rho(\tau)\der{x^\sigma}\Phi(x(\tau)) ~~\rightarrow~~ F_{ij}(x(\tau_0))=\eps_{ijk}\der{x^k}R \Phi(x(\tau_0))~.
\end{equation}
As the Higgs-fields for monopoles $\Phi_{\rm m}$ and self-dual strings $\Phi_{\rm sds}$ are related via $\Phi_{\rm m}=R\Phi_{\rm sds}$, we recover the Bogomolny monopole equation.

\subsection{Comment on non-abelian self-dual strings}

One can easily extend the loop space version of the self-dual string equation to the situation in which gauge and Higgs fields take values in the adjoint representation of $\au(N)$:
\begin{equation}\label{eq:nonabelian}
 F_{\mu\nu}(x(\tau)):=[\nabla_{\mu},\nabla_\nu]=\eps_{\mu\nu\rho\sigma}\xd^\rho(\tau) \nabla_\sigma\Phi(x(\tau))\ewith\nabla_\mu:=\der{x^\mu}+A_\mu(x(\tau))~.
\end{equation}
One readily verifies that the fields \eqref{eq:MFieldDefinitions}, when constructed from appropriate matrices of zero modes $\psi_{s,x(\tau)}$, satisfy \eqref{eq:nonabelian}. Also the reduction procedure given in the previous section works in the non-abelian case. We therefore postulate that the non-abelian version of the loop space self-dual string equation \eqref{eq:nonabelian} is the appropriate description of a stack of M2-branes ending on $N$ M5-branes. It might be possible to translate this equation back to $S^3$ and thus to $\FR^4\supset S^3$ via a regression and one might thus approach an effective field theory for stacks of M5-branes. All this should be put within the context of non-abelian gerbes and further consistency checks from a physical perspective are in order, too. This, however, is beyond our scope here and we leave it to an upcoming paper.

\section{Discussion and future directions}

We developed an ADHMN-like construction of solutions to the world-volume theory of M5-branes corresponding to self-dual strings. This construction contains all the ingredients of the usual ADHMN construction: One derives a gauge potential and a Higgs field from the normalizable zero modes of a Dirac operator which is constructed from solutions to certain BPS equations.

The two-form $B$-field of the self-dual string belongs to a connective structure of a gerbe over $S^3\subset\FR^4$, which can be mapped via a transgression to a connection on a principal fibre bundle over the loop space $\CL S^3$. We could therefore perform our construction on $\CL S^3$ by translating the self-dual string equation to this loop space. The resulting equation is an ordinary gauge field equation and can therefore be rendered non-abelian. We suggested that this is in fact a suitable BPS equation for an effective description of stacks of multiple M5-branes.

We demonstrated how our construction is linked to the ordinary ADHMN construction by a reduction process and T-duality. We also constructed explicit self-dual string solutions on $\CL S^3$ using our algorithm and verified that the inverse of the transgression map takes them back to self-dual strings on $S^3\subset \FR^4$.

It should be stressed that we therefore agree with the conclusion of \cite{Gustavsson:2007vu}: The ability of giving an extension of the ADHMN construction for self-dual strings (together with the lack of sufficiently many finite-dimensional 3-Lie algebras to describe stacks of arbitrarily many M2-branes) suggests that one should seek descriptions involving loop spaces, or even better, gerbes.

There is now a wealth of directions for future study. First of all, the non-abelian extension of our ADHMN construction should be interpreted in the context of non-abelian gerbes and one should try to make sense out of the non-abelian equation on $\CL S^3$ in terms of fields on $S^3$. Many details remain to be worked out in this case, as e.g.\ the exact boundary conditions and the reduction of the number of zero modes of the Dirac operator. Second, a full generalization of the Nahm transform, possibly working on special loop spaces of $T^5$, should be developed. Simpler aims are to extend our construction to the ABJM case\footnote{The corresponding Basu-Harvey equation was given in \cite{Terashima:2008sy,Hanaki:2008cu}.}, the case of the $\CN=2$ models constructed in \cite{Cherkis:2008qr} as well as to the case of the higher Nahm-type equations discussed in \cite{Lazaroiu:2009wz}. Studying noncommutative deformations in this context would also be interesting. In the long term, it should be possible to mimic the link of the ADHMN methods via monads to the twistor formalism (see e.g.\ \cite{Hitchin:1983ay}) also for our construction. This would lead to twistor spaces for the description of self-dual strings.

The fact that our construction fits nicely into both the mathematical and physical contexts is exciting. It is therefore reasonable to hope that following the lines of research proposed above will help to shed new light on the effective description of M5-branes.

\acknowledgements
I would like to thank Werner Nahm for encouraging comments on ideas leading to this work. I also thank Sam Palmer and Martin Wolf for helpful comments. I am particularly grateful to Sergey Cherkis for many helpful and enjoyable discussions on M2-branes in the past and for detailed comments on a first draft of this paper. This work was supported by a Career Acceleration Fellowship from the UK Engineering and Physical Sciences Research Council.

\appendices

\subsection{3-Lie algebras and representations}

A {\em metric 3-Lie algebra} is an inner product space $(\CA,(\cdot,\cdot))$ endowed with a trilinear, totally antisymmetric bracket $[\cdot,\cdot,\cdot]:\CA^{\wedge 3}\rightarrow\CA$, which satisfies the {\em fundamental identity} \cite{Filippov:1985aa}
\begin{equation}
 [a,b,[c,d,e]]=[[a,b,c],d,e]+[c,[a,b,d],e]+[c,d,[a,b,e]]
\end{equation}
and which is compatible with the inner product
\begin{equation}
 ([a,b,c],d)+(c,[a,b,d])=0
\end{equation}
for all $a,b,c,d,e\in\CA$. The {\em inner derivations} ${\rm Der}(\CA)$ are the linear extensions of the maps $D(a, b):=[a,b,\cdot\,]$ with $a,b,\in\CA$. Due to the fundamental identity, they form a Lie algebra, the {\em associated Lie algebra} $\frg_\CA$,
\begin{equation}
\begin{aligned}
{}[D(a, b),D(c, d)]\acton e&=[a,b,[c,d,e]]-[c,d,[a,b,e]]=[[a,b,c],d,e]+[c,[a,b,d],e]\\&
=(D([a,b,c], d)+D(c, [a,b,d]))\acton e~.
\end{aligned}
\end{equation}

The most prominent example of a 3-Lie algebra is $A_4$, which corresponds to $\FR^4$ with standard basis $e_\alpha$, $\alpha=1,\ldots,4$, together with the 3-bracket
\begin{equation}
 [e_\alpha,e_\beta,e_\gamma]=\eps_{\alpha\beta\gamma\delta}e_\delta~.
\end{equation}
The euclidean scalar product, $(e_\alpha,e_\beta)=\delta_{\alpha\beta}$, is compatible with this 3-bracket and thus $A_4$ is a metric 3-Lie algebra. Its associated Lie algebra is $\frg_{A_4} =\asu(2)\times \asu(2)$. For an exhaustive review on $n$-ary Lie algebras, see \cite{deAzcarraga:2010mr}.

In our discussion, we need a representation $\rho$ of the associated Lie algebra compatible with the 3-Lie algebra structure in the sense that
\begin{equation}
 D^{(\rho)}(a,b)\cdot D^{(\rho)}(c,d)-D^{(\rho)}(c,d)\cdot D^{(\rho)}(a,b)=D^{(\rho)}([a,b,c],d)+D^{(\rho)}(c,[a,b,d])~,
\end{equation}
where $D^{(\rho)}(a,b)$ denotes the inner derivation $D(a,b)$ in the representation $\rho$. As an example of such a representation, consider the spinor representation $\rho$ of $\frg_{A_4}$ on $\FC^4$,
\begin{equation}
 D^{(\rho)}(e_\alpha,e_\beta)\zeta:=\tfrac{1}{2}\gamma^{\alpha\beta}\gamma_5 \zeta~,
\end{equation}
where $\zeta \in \FC^4$ and $\gamma^{\alpha\beta}=\tfrac{1}{2}[\gamma^\alpha,\gamma^\beta]$ with $\gamma^\alpha$ being the generators of the Clifford algebra $Cl(\FR^4)$ satisfying $\{\gamma^\alpha,\gamma^\beta\}=+2\delta^{\alpha\beta}$. Using this representation, we also find the 3-bracket of $A_4$ as described in \cite{Gustavsson:2007vu}:
\begin{equation}
 [e_\alpha,e_\beta,e_\gamma]=[D^{(\rho)}(e_\alpha,e_\beta),\gamma_\gamma]=\eps_{\alpha\beta\gamma\delta}e_\delta~.
\end{equation}



\end{document}